\documentstyle[12pt]{article}

\newcommand{\be}{\begin{equation}}
\newcommand{\ee}{\end{equation}}
\newcommand{\bea}{\begin{eqnarray}}
\newcommand{\eea}{\end{eqnarray}}

\newcommand{\I} {{\cal I}}
\newcommand{\r} {r_{+}}
\newcommand{\re} {(r_{+})_{ext}}

\begin{document}

\begin{center}
\begin{large}
{\bf  Radiation via Tunneling \\}
{\bf in the \\}
{\bf Charged BTZ Black Hole \\}
\end{large}  
\end{center}
\vspace*{0.50cm}
\begin{center}
{\sl by\\}
\vspace*{1.00cm}
{\bf A.J.M. Medved\\}
\vspace*{1.00cm}
{\sl
Department of Physics and Theoretical Physics Institute\\
University of Alberta\\
Edmonton, Canada T6G-2J1\\
{[e-mail: amedved@phys.ualberta.ca]}}\\
\end{center}
\bigskip\noindent
\begin{center}
\begin{large}
{\bf
ABSTRACT
}
\end{large}
\end{center}
\vspace*{0.50cm}
\par
\noindent

It is often convenient to view   Hawking (black hole)
radiation  as a process of quantum tunneling.
Within this framework, Kraus and Wilczek (KW)  have initiated an
analytical treatment of black hole emission. Notably,
their  methodology incorporates the effects of a dynamical
black hole geometry. In the current paper, the
KW formalism is applied to the case of a charged BTZ
black hole. In the context of this interesting model,
we are able to demonstrate a non-thermal spectrum,
with  the usual Hawking result being reproduced
at  zeroth order in frequency.
Considerable attention is then given to the examination
of near-extremal thermodynamics.

\newpage

\section{Introduction}
\par
Roughly three decades ago, Bekenstein and Hawking established the 
well-known  analogy that exists between black hole mechanics 
and thermodynamic
systems \cite {bek,haw}. This intriguing relationship, which
appears to have even deeper physical significance,  has
since been the subject  of a ``countless'' number of investigations. However,
in spite of all this attention, black hole
thermodynamics continues to  have several  unresolved issues.
For example, the microscopic origin of the Bekenstein-Hawking
 entropy is currently a prominent
open question.
(See Ref.\cite{wald} for a review and references.) 
\par
Although better understood than the origin of  entropy, the mechanism of
 black hole radiance
remains shrouded in some degree of mystery. To review,
it was Hawking who first  demonstrated that black holes radiate
(via quantum effects) such that the emission spectrum (at infinity)
is essentially thermal \cite{haw2}. It was this
remarkable discovery  that gave physical credibility
to the thermodynamic analogy; but at a cost, as this result has
dramatic implications regarding our understanding of
quantum evolution. More specifically, the  evaporation 
of a black hole in this manner implies that a pure state (the original matter
that forms the black hole) can evolve into a mixed state (the
thermal spectrum at infinity). Such an evolution is a violation 
of the fundamental principles of quantum theory, as these prescribe
a unitary time  evolution of  basis states.     
This contradictory nature of black hole radiation has been often  labelled 
as the  ``information loss
paradox'' \cite{haw3}.
\par
The above paradox can perhaps be attributed to the semi-classical nature
of such investigations, as a conspicuously absent quantum  theory of
 gravity remains
a formidable obstacle. However, there is  another  fundamental
issue that must necessarily be dealt with; 
namely, energy conservation.  It seems clear that a radiating
black hole should be losing mass (which is directly related
to the temperature), but this dynamical effect is often neglected
in formal treatments. 
\par
To further explore the issue of dynamics, it proves convenient 
to adapt the viewpoint (commonly accepted, but often overlooked) 
that Hawking radiation is really  a quantum tunneling
process. According to this scenario, a pair of particles
is spontaneously created just inside of the black hole horizon.
One of the particles
then ``tunnels'' out to the opposite side, where it emerges
with positive energy. Meanwhile, the negative-energy ``partner''
remains behind and effectively lowers the mass of the black hole.
\par
In a program of study that was initiated by Kraus and Wilczek (KW) 
\cite{t1,t2,t3}
(since developed and  generalized 
in 
Refs.\cite{t4,t5,t6,t7,t8,t9,t10}\footnote{Also of interest is Ref.\cite{new1},
where the tunneling picture has been applied via a different methodology.}), 
this tunneling picture was the 
foundation
for a dynamical treatment of black hole 
radiance.
In a nutshell,  KW considered the effects of a positive-energy
matter shell (i.e., ``s-wave'') propagating outwards through
the horizon of a spherically symmetric black hole.
The pertinent point of their work was the incorporation
of a  dynamical description of the black hole background.
More specifically, the background geometry is allowed to fluctuate
and thus support a black hole of varying mass. (That is, the
``self-gravitation'' of the radiation is taken into account.)
Such formalism  leads to the
enforcement of
energy conservation in a natural way. In particular, KW 
allow the black hole to lose mass while
radiating, but maintain a constant energy for the total system.
\par
Another salient point of the KW method was their choice of boundary
conditions on the  quantum matter fields. These
are effectively enforced  via the choice of  coordinates that
have been
used to foliate  the spacetime. In their analysis, KW implemented
so-called  ``Painleve''  coordinates\footnote{The coordinate system in
question  was first proposed by Painleve in 1921 \cite{pan}.} 
  that are not only time independent and regular at the horizon,
 but for which  time reversal
is manifestly  asymmetric (unlike the often-studied Kruskal coordinates 
\cite{kru}). 
That is, the coordinates are stationary
but not static.  This  gauge choice seems to  be particularly
appropriate for describing   the geometry of a slowly evaporating
 black hole.
\par
Let us summarize  some important findings of the KW analysis, which
 will be exploited later in the paper. Firstly, a  canonical  Hamiltonian
formulation  (in the previously discussed framework) yields
the remarkably simple result for the total action of the system:
\be
\I=\int d\tau \left[{dr\over d\tau}p_r  + p_{\tau}\right].
\label{0}
\ee
Here, $r$ and $\tau$ are respectively the radial and temporal
coordinate in the Painleve gauge, while $p_{\mu}$  represents
the  conjugate momentum of coordinate $\mu$. 
Secondly,  a semi-classical (WKB) approximation\footnote{It should
be kept in mind that such an approximation treats the  radiating matter
as point particles. Hence, it is only appropriate when
in a regime of  sufficiently large black holes.}
leads to the following expression for the emission rate ($\Gamma$):
\be
-{1\over 2}\ln(\Gamma) \approx  Im \I,
\label{00}
\ee 
where only the first term in Eq.(\ref{0}) contributes
to the imaginary part of the action.
With these expressions, it can be shown  that the spectrum of black hole
radiation is not strictly thermal; rather, it contains a
frequency-dependent ``greybody'' factor. (Although this greybody factor
is well known \cite{haw2}, it is often neglected in the relevant literature.) 
\par
The purpose of the current paper is to examine
the thermodynamics of a charged BTZ black hole via a method
based on the KW formalism.
The BTZ theory refers to special solutions
of 2+1-dimensional anti-de Sitter gravity having  all of the properties
of black holes. (These solutions were first identified
by Banados, Teitelboim and Zanelli \cite{btz1}.)
 Charged BTZ black holes are simply the analogous
solutions in 2+1-dimensional AdS-Maxwell gravity \cite{btz2,btz3}. 
Although a ``toy'' model in some respect, the BTZ black hole has
stirred significant interest by virtue of its connections 
with certain string theories \cite{btzs}
 and  its role in microscopic entropy
calculations \cite{car}. Furthermore, the BTZ model has proven
to be an especially  useful ``laboratory'' for studying  quantum-corrected
thermodynamics \cite{new2,new3,new4,new5,new6,new7}.
\par
This paper is organized as follows. In Section 2, we begin by
introducing the model of interest  and the solution
corresponding to a charged BTZ black hole. This is 
followed by a semi-classical calculation of the black
hole emission rate from which the spectrum can be directly extrapolated.
In Section 3, we elaborate on our results; particularly,
in the context of near-extremal black holes.
It is worth noting that the KW formalism has particular
importance near the extremal limit, where even the smallest changes
in the black hole mass can  significantly deform  the background geometry.
Section 4 ends  with a brief summary and concluding
remarks.

\section{Analysis}
\par
The model of interest, 2+1-dimensional AdS-Maxwell gravity, can
be described by the following gravitational action:
\be
\I_{G}={1\over 4}\int d^3 x \sqrt{-g}\left[ {1\over 4\pi G}
(R-2\Lambda)-F^{\mu\nu}F_{\mu\nu}\right].
\label{1}
\ee
Here, $G$ is the 3-dimensional Newton  constant 
(with dimensions of inverse mass) and $\Lambda=-l^{-2}$
is the negative cosmological constant. Note that we
are assuming vanishing rotation for the sake of
simplicity.
\par
A static, charged black hole solution has been found
for the above action \cite{btz1,btz2,btz3}.
This can be expressed as follows:
\be
ds^2=-N^2(r)dt^2+N^{-2}(r)dr^2+r^2 d\phi^2,
\label{2}
\ee
where:
\be
N^2(r)={r^2\over l^2}-8GM-Q^2\ln\left({r^2\over l^2}\right).
\label{3}
\ee
Here, $M$ is the generalized ADM mass \cite{adm} of the charged
 BTZ black hole
 and $Q$ is a dimensionless parameter that
represents the  charge.
\par
It will often prove convenient to re-express Eq.(\ref{3})
in the following form:
\be
N^2(r)={r^2\over l^2}-{\r^2\over l^2}-Q^2\ln\left({r^2\over \r^2}\right),
\label{4}
\ee
where   $\r$ (the black hole horizon) is the outermost value of $r$  for which
$N^2(\r)=0$. Typically, there will exist some second value $r_{-}\leq \r$
such that $N^2(r_{-})=0$ as well.
\par
To obtain the condition of extremality, $r_{-}=\r$, let us consider
the Hawking temperature ($T_H$) as determined by the surface
gravity ($\kappa$) at the horizon \cite{gh}:
\be
T_{H}={\kappa\over 2\pi}={1\over 4\pi}\left.{dN^2\over dr}\right|_{r=\r}
= {\r\over 2\pi l^2}\left(1-{ l^2 Q^2 \over \r^2}\right).
\label{5}
\ee
It follows that the black hole becomes extremal when $T_H$ vanishes
or at:
\be
\re^2= l^2 Q^2.
\label{6}
\ee
\par
With black hole emission in mind,
we now  begin a  semi-classical calculation
 that is based on the  Kraus-Wilczek treatment
\cite{t1,t2,t3}. (Also see Refs.\cite {t4}-\cite{t9}.)
 First, it is appropriate
to re-express the metric in a form that is manifestly stationary
(but not static) and regular at the horizon. These ``Painleve-like''
coordinates \cite{pan} can be obtained with the following 
redefinition of the time coordinate:
\be
d\tau =dt -{\sqrt{1-N^2(r)}\over N^2(r)} dr.
\label{7}
\ee
Substituting into Eq.(\ref{2}), we now have:
\be
ds^2=-N^2(r)d\tau^2+dr^2+2\sqrt{1-N^2(r)}d\tau dr+r^2 d\phi^2.
\label{8}
\ee 
\par
It is useful to evaluate the radial, null
geodesics. Under these conditions 
($d\phi = ds^2 = 0$),
 Eq.(\ref{8}) reduces to:
\be
0=-N^2(r) +{\dot r}^2+ 2\sqrt{1-N^2(r)}{\dot r},
\label{9}
\ee
where ${\dot r}=dr/d\tau$. Solving for ${\dot r}$, we find:
\be
{\dot r}=\pm 1 -\sqrt{1-N^2(r)},
\label{10}
\ee
where the $+/-$ sign can be identified with outgoing/incoming
radial motion.
\par
Next, we consider a self-gravitating shell of
positive energy ($\omega$) radiating outwards through the
black hole horizon. For simplicity,  we assume a shell having
 zero rest mass, zero charge and symmetry  with respect to the
angular coordinate. Our viewpoint will be that the total
mass of the system stays fixed, while the black hole mass
varies according to $M\rightarrow M-\omega$. It then follows
that the shell of energy travels along geodesics 
which are described by:
\be
{\dot r}= 1 -\sqrt{1-N^2(r,M-\omega)}.
\label{11}
\ee
\par
In our analysis, the primary interest is the semi-classical emission
rate of this shell-charged BTZ system. That is (cf. Eq.(\ref{00})):
\be
\Gamma(\omega) = e^{-\omega/T(\omega)}\approx e^{-2 Im\I},
\label{12}
\ee
where $T(\omega)$ is the temperature at
 ``frequency'' $\omega$.
\par 
For an  positive-energy ``s-wave'' propagating outwards,
KW have shown 
that the imaginary part of the total action
can be expressed 
as \cite{t2}:\footnote{Although the original
KW analysis was for a spherically symmetric system
in an asymptotically flat spacetime, this  formalism
has since been generalized for AdS spacetimes with an arbitrary
number of dimensions \cite{t7}.}         
\be
Im\I =Im\int d\tau {\dot r}
 p_r =Im\int^{r_{out}}_{r_{in}}\int^{p_r}_{0}
dp^{\prime}_r dr,
\label{13}
\ee 
where $p_r$ is the canonical momentum (conjugate to $r$)
and the total action includes the gravitational action ($\I_G$)
along with the action for the shell.
\par
At this point,
it is useful to apply Hamilton's equation: ${\dot r}=dH/dp_r=
d(M-\omega)/dp_r$. Hence, Eq.(\ref{13}) can be re-written
in the following manner:
\bea
Im\I &=& Im\int^{r_{out}}_{r_{in}}\int^{\omega}_{0}
{-d\omega^{\prime}dr\over {\dot r}(r,M-\omega^{\prime})}
\nonumber \\
&=&
Im\int^{\r(M-\omega)}_{\r(M)}\int^{\omega}_{0}
{-d\omega^{\prime}dr\over 
1-\sqrt{1-{r^2\over l^2}+8G\left[M-\omega^{\prime}\right]+Q^2\ln\left(
{r^2\over l^2}\right)}},
\label{14}
\eea
where we have applied Eqs.(\ref{3},\ref{11}) in attaining the lower line.
\par
The integration over $\omega^{\prime}$ can be readily done as
a contour integral. Significantly to this calculation, the correct sign
is obtained via the requirement that $\omega\rightarrow \omega
-i\delta$ (where $\delta >0$). This choice ensures that the
positive-energy solution ($\sim e^{-i\omega\tau}$) decays
in time.  
\par
For the explicit evaluation of this integral,  let us
first change variables, $\mu= 
1-{r^2\over l^2}+8G\left[M-\omega\right]+Q^2\ln\left(
{r^2\over l^2}\right)$,
 to obtain the following form:
\be
Im\I=Im\int^{\r(M-\omega)}_{\r(M)}\int^{\mu(\omega)}_{\mu(0)}
{d\mu^{\prime}\over 1-\sqrt{\mu^{\prime}}} {dr\over 8G}.
\ee
Given that the above condition (on $\omega$) 
implies $\mu\rightarrow\mu+i{\tilde\delta}$
(where ${\tilde\delta} > 0$) and that
 $\mu(\omega) < \mu(0)$, it is appropriate to
integrate clockwise in the upper half of the complex-$\mu^{\prime}$ plane.
This process yields:  
\bea
Im\I &=& -{\pi\over 4G}\int^{\r(M-\omega)}_{\r(M)}dr
\nonumber \\
&=& {\pi\over 4G}\left[\r(M)-\r(M-\omega)\right].
\label{15}
\eea
\par
We will  now proceed to evaluate the quantity $\r(M-\omega)$ via
the following argument.
First, there must exist some real function $\eta=\eta(\omega)$
such that $\r^2(M)\rightarrow \r^2(M)-\eta^2$ as $M\rightarrow M-\omega$.
A comparison of Eqs.(\ref{3},\ref{4}) then yields the relation:
\be
{\r^2(M)-\eta^2\over l^2}- 8G\left[M-\omega\right]-Q^2\ln\left(
{\r^2(M)-\eta^2\over l^2}\right)=0.
\label{17}
\ee
Eliminating the zeroth-order terms, we have:
\be
8G\omega={\eta^2\over l^2}+Q^2 \ln\left(
1-{\eta^2\over \r^2(M)}\right).
\label{18}
\ee
\par
For a sufficiently large black hole (as is appropriate for 
a semi-classical analysis), it follows that $\eta^2<<\r^2(M)$.
Hence, we can expand the above logarithm to obtain:
\be
\eta^2\approx {8Gl^2\r^2(M)\over \r^2(M)-l^2Q^2}\omega.
\label{19}
\ee
\par
Recalling Eq.(\ref{6}) for the extremal limit ($\re^2=l^2Q^2$),
we have  an apparent breakdown in the formalism when this limit is approached.
This is not surprising, as  black hole geometries
are known to be altered dramatically in this limiting case \cite{teit,ext1}.
To examine this issue more carefully, let us consider the next
term in the logarithmic expansion.  This addition results
in a quadratic expression for $\eta^2$, which can be solved to
yield:
\be
\eta^2={\r^2(M)-l^2Q^2\over l^2Q^2}\pm\sqrt{
{\left[\r^2(M)-l^2Q^2\right]^2\over l^4Q^4}
-{16G\r^2\over Q^2}\omega}.
\label{19.5}
\ee
Since $\omega$ is  presumed to be  positive,
the square root tends to an imaginary quantity in the
extremal limit (unless $\omega=0$). Thus, the  extremal breakdown
in the formalism appears to persist.
 The problem may be linked to the prior assumption
of $\eta^2<<\r^2(M)$. It is possible that this is not a valid
constraint when probing the extremal condition. 
We conjecture, however,  that it will always be possible to
choose a sufficiently large enough black hole so that this condition
is valid even  in a near-extremal regime. In this case,  
it follows from Eq.(\ref{19.5}) that $\omega$ (and, hence, $\eta^2$)
 goes rapidly
to zero as the extremal limit is approached.
The obvious implication is that
 a (sufficiently large) black hole will cease to
radiate as it approaches extremality (which also follows
from the third law of thermodynamics). This point
will be elaborated on in Section 3; until then, we focus
on   black holes far from extremality.
\par
Incorporating the  result of the above analysis (\ref{19}) into Eq.(\ref{15}),
we obtain:
\be
Im\I 
\approx {\pi\over 4G}\left[\r(M)-\sqrt{\r^2(M)- 
{8Gl^2\r^2(M)\over \r^2(M)-l^2Q^2}\omega}
\right].
\label{20}
\ee
The square root can be expanded to yield:
\be
Im\I\approx \omega\left[{\pi l^2 \r(M) \over \r^2(M)-l^2Q^2} 
-{2\pi G l^4 \r(M)\over \left[\r^2(M)-l^2 Q^2\right]^2}\omega 
+ \quad  ... \quad \right],
\label{21}
\ee
where ``...''  represents the higher-order (in $\omega$)
corrections.
\par
By recalling from Eq.(\ref{12}) that $2Im\I\approx \omega / T(\omega)$,
we are now able to deduce the  black hole temperature
at any  given frequency:
\be
T(\omega)\approx {\r^2(M)-l^2Q^2\over 2\pi l^2 \r(M)}\left[
 1+ {2Gl^2 \over \r^2(M)-l^2Q^2}\omega + \quad ... \quad \right],
\label{22}
\ee
where the higher-order terms (...) are essentially an expansion in
powers of $\omega/(\r^2(M)-l^2Q^2)$. 
Notably, the zeroth-order  terms
reproduce  the expected value for the Hawking temperature; cf. Eq.(\ref{5}).
However, the higher-order quantum corrections, which are 
 clearly non-vanishing,
lead to a frequency-dependent
``greybody'' factor. That is, the  emission spectrum
deviates from that of a pure black body. Strictly speaking,
the black hole emits non-thermal radiation !
\par
A useful check on this formalism follows from the first
law of thermodynamics.
 Consider that, during black hole emission, the expected change in entropy
is given by $\Delta S=\Delta M/T=-\omega/T$. That is (cf. Eq.(\ref{12})):
\be
\Delta S = -2 Im\I.
\label{23}
\ee
An inspection of Eq.(\ref{15}) thus yields:
\be
S(M)={\pi\r(M)\over 2G}\quad + \quad constant.
\label{24}
\ee 
For a vanishing constant,
this is just the Bekenstein-Hawking area law \cite{bek,haw}
of $S=A_{+}/4G$
 (generalized
to a 3-dimensional black hole).

\section{Discussion}
\par
It is instructive to consider the following observation:
the black hole emission rate,  $\Gamma\approx e^{-2Im\I}$,
is a measurable and (hence) real quantity. This
restricts the square root in Eq.(\ref{20}) 
to be a real quantity as well.  Thus, naively, it follows that:
\be
\r^2(M) -\re^2 \geq 8Gl^2\omega.
\label{25}
\ee
We say ``naively'' because of the formal  breakdown that
occured in the extremal limit. (See the prior section.) However, our 
conjecture  that constrains  the horizon geometry
of  sufficiently large black holes, $\eta^2<<\r^2(M)$,
effectively implies  the same condition. (Again, see  last section.)
So, at least  on the basis of these arguments,  
radiation past extremality is impossible; thus
 demonstrating a  natural enforcement
of the third law of thermodynamics.
To put it another way:
if  the black hole ceases to radiate  in the extremal limit, then the classical
result of  vanishing extremal temperature remains essentially valid, and
(hence) the  emission rate vanishes exponentially as this limit is approached.
That is,  there is  zero possibility of
a black hole decaying into a naked
singularity !
\par
Although naked singularities are censored against, one
might still wonder if a state of absolute
extremality can be achieved.
 This is  a difficult  dilemma to resolve, one way or
the other. However, studies elsewhere in the literature
have argued that extremal and non-extremal black 
holes are qualitatively distinct objects
\cite{teit,ext1,ext2}.
Such arguments are primarily based on the topological
differences that exist between  extremal and non-extremal
spacetimes.  Moreover, these  differences seem to imply
that a non-extremal black hole would {\bf not} be
able to continuously deform into an extremal one
(and {\it vice versa}).\footnote{There is, however,
a wealth of literature that has argued (directly or
indirectly) on behalf of
a viable extremal limit. The most compelling of these being
the microscopic  calculations of extremal entropy
in the context of string theory \cite{str}.}
Also of note,  this viewpoint has been substantiated
by a recent  investigation into the physical  spectra of  charged 
black holes \cite{new8}: generically,  extremal black holes 
can {\bf not} be 
achieved (at the quantum level) due to vacuum fluctuations in the 
horizon.
\par
For the sake of argument, let us
accept the above conjecture and also accept
that a black hole can  not come arbitrarily close
to an extremal state.\footnote{This latter assertion
follows directly from the above mentioned study on physical
spectra \cite{new8}. For further arguments along this line, 
see Ref.\cite{blah}.}
  Furthermore,  let us 
 assume that the black hole ``freezes''
at a point suitably far from extremality. In this case (and with charge
regarded as a fixed quantity), 
the  prior formalism can be shown to
yield  
   lower bounds on the black hole
 temperature and entropy.  We demonstrate this as follows.
\par 
 First, if the smallest allowed
quanta of energy is taken to be $\epsilon$,\footnote{One might expect
such quanta to be roughly on the order of the Planck mass.} then
Eq.(\ref{25}) leads to the following near-extremal (``ne'') limit for
the horizon radius:
\be
(\r)_{ne}^2 \approx l^2Q^2 +8Gl^2 \epsilon.
\label{26} 
\ee
By substituting this relation into Eq.(\ref{22}),
we can  obtain  a near-extremal
bound on the temperature.\footnote{It is interesting to
note that a similar bound was arrived at by a much different rationalization
in Ref.\cite{blah}.} To first order in $\epsilon$, the following is found:
\be
T_{ne}\approx {4G \over \pi l \sqrt{Q^2}} \epsilon.
\label{27}
\ee
Similarly, we can obtain a near-extremal  bound on the entropy:
\be
S_{ne}\approx S_{ext} + {2\pi \over \sqrt{Q^2}} \epsilon,
\label{28}
\ee
where $S_{ext}=\pi\sqrt{l^2 Q^2}/2G$.
\par
Let us take note of a related work that has 
recently been carried out by Vagenas.\footnote{Vagenas applied
the KW program  to a ``string-inspired'' (charged) dilaton theory \cite{t9}
and a rotating (but not charged) BTZ model \cite{t10}.} The viewpoint of
this study was the existence of a well-defined 
extremal limit. On this basis, Vagenas has
proposed that extremal thermodynamics  are calculated
by: (i) fixing $\omega$  such that the condition of extremality
($r_{-}=r_{+}$) is satisfied for $M-\omega$ (rather than $M$)
and then (ii) using this extremal  constraint to eliminate $\omega$ from
the thermodynamic expressions.  
\par
If we fix $\omega$ (or, equivalently, $\eta$)  in an analogous fashion, then
the following extremal constraint can be obtained:
\be
\eta_{ext}^2=\re^2-l^2 Q^2,
\label{28.5}
\ee
where $\re$ is now the ``revised'' extremal horizon.  Substituting
this constraining relation into Eq.(\ref{17}), we find:
\be
\omega_{ext}=M+{Q^2\over 8G}\left[\ln\left({Q^2\over l^2}\right)-1
\right]. 
\label{29}
\ee
This result implies that $\omega_{ext}$ will be at least on the
order of $M$, which directly  contradicts 
our previous observation (applicable to sufficiently large black holes):  
$\omega\rightarrow 0$
as the extremal limit is approached.
\par
The  contradictory behavior  of the extremal limiting
case may be a consequence of an apparent
pathology in  the charged BTZ black hole. (This pathology
is described in Section 4.) 
However, this extremal breakdown may rather be a manifestation of
the third law of thermodynamics; that is, an extremal
limit of a non-extremal calculation may actually  be erroneous procedure.
In the latter case, Eqs.(\ref{27},\ref{28}) can  be regarded,
 at least in some schematic sense,
as the correct near-extremal bounds on the temperature and entropy.

\section{Conclusion}
\par
 In the preceding paper, we have considered the thermal emission
from a charged (non-rotating) BTZ black hole. Our analysis
followed the viewpoint that Hawking radiation is due
to a tunneling process and was appropriately based on the 
methodology of Kraus and Wilczek \cite{t1,t2,t3}. The pertinent point of this
approach is that black hole radiance is a dynamical
mechanism for which
energy conservation must be enforced.
\par
A rigorous application of the prescribed method allowed us to
verify  the ``standard'' results:
the Hawking temperature  (at  zeroth order in frequency)
 and the Bekenstein-Hawking
entropy \cite{bek,haw,haw2}.
Furthermore, the black hole temperature was found to have
frequency-dependent corrections (i.e., a greybody factor), which means that
the emission spectrum actually deviates from thermality. 
Often, the relevant literature conveniently depicts the black hole spectrum
as being perfectly thermal, although it is well known
that this is not actually the case \cite{haw2}. 
\par
 In general, one can
evaluate such greybody factors (at a semi-classical level)
by solving the appropriate Klein-Gordon equation
on a fixed black hole background \cite{ms}.
 Notably, such a calculation has been considered for a rotating BTZ black hole
\cite{bss,lm}.  Although it would be interesting to
compare these results with our  derivations,  this seems impratical
given 
 the complexity of their expressions and
the lack of  compatibilty with the charged BTZ scenario. 
\par
We also considered the case of black hole extremality
and found that naked singularities are forbidden 
by this formalism in a natural way.
Using  the premise that extremal and non-extremal
black holes are qualitatively distinct entities \cite{teit,ext1}
(and other conjectural considerations),
we were able to evaluate ``near-extremal'' limits
to the black hole temperature and entropy. Meanwhile,
the alternate viewpoint (of a well-defined extremal limit)
gave rise to an apparent contradiction.
\par
Before concluding, a couple of points are in order.
Firstly, it has been suggested that the charged BTZ black hole
is a somewhat pathological model \cite{btz2}. The reasoning
is as follows: (i) it exists for arbitrarily negative
values of mass and (ii) there is no upper bound on
the electric charge. (Such behavior is  contrary to,
for instance, that of the  Reissner-Nordstrom black hole.)
We point out, however, that a charged black hole
will tend to discharge by preferentially
emitting charged particles of the same sign as
itself \cite{bd}. Hence, black hole emission may serve
as a natural mechanism for effectively suppressing the charge
(relative to the mass). Perhaps, such a mechanism would be
analogous to that which censors naked singularities.
This question could possibly be addressed in a more realistic (but
more complex) study.
\par
Secondly, we again remind the reader that the preceding study
was a semi-classical analysis. In fact, the formal methods of
Kraus and Wilczek are akin to a WKB approximation, meaning that
the radiation should be treated  as point particles.
Such an approximation can only be valid in a regime
of sufficiently  large black holes. If we are
to properly address the thermodynamics of
microscopic black holes,  then a better understanding of  physics
at the Planck scale  is a necessary prerequisite.

\section{Acknowledgments}
\par
The author  would like to thank  V.P.  Frolov  for helpful
conversations. 
  \par\vspace*{20pt}


\end{document}